\documentclass[showpacs,preprintnumbers,amsmath,amssymb,superscriptaddress]{revtex4}
\usepackage{graphicx}
\usepackage{aas_macros}

\input{colordvi.tex}

\begin{document}
%
\title{Constraint on the primordial vector mode and its magnetic
field generation from seven-year Wilkinson Microwave Anisotropy Probe
Observations}
\author{Kiyotomo Ichiki}
\email{ichiki@a.phys.nagoya-u.ac.jp}
\affiliation{%
Department of Physics and Astrophysics, Nagoya University, Nagoya
464-8602, Japan
}
\author{Keitaro Takahashi}
\affiliation{%
Department of Physics, Kumamoto University, Kumamoto
860-8555, Japan
}
\author{Naoshi Sugiyama} 
\affiliation{%
Department of Physics and Astrophysics, Nagoya University,
Nagoya 464-8602, Japan
} 
\affiliation{%
Kobayashi-Maskawa Institute for the Origin of Particles and
the Universe, Nagoya University, Chikusa-ku, Nagoya, 464-8602, Japan
}
\affiliation{%
Institute for the Physics and Mathematics of the Universe (IPMU), 
The University of Tokyo, Chiba 277-8582, Japan
}%

\date{\today} \preprint{}
\begin{abstract}
 A primordial vector mode and its associated magnetic field generation
 are investigated.  Firstly, we put an observational constraint on the
 amount of the primordial vector mode from the seven-year WMAP data.
 The constraint is found as $r_v \lesssim -\frac{r}{40}+0.012$, where
 $r_v$ and $r$ are the amounts of vector and tensor perturbation
 amplitudes with respect to the scalar one, respectively. Secondly, we
 calculate the spectrum of magnetic fields inevitably created from the
 primordial vector mode, given the constraint on $r_v$. It is found that
 the maximum amount of magnetic fields generated from the vector
 mode is given by $B\lesssim 10^{-22}{\rm G}
 \left(\frac{r_v}{0.012}\right)^{1/2}
 \left(\frac{k}{0.002}\right)^{(n_v+1)/2}$ with $n_v$ being a spectral
 index of the vector mode.  We find a non-trivial
 cancellation of the magnetic field generation in the radiation
 dominated era, which creates a characteristic cut off in the magnetic
 field spectrum around $k\approx 1.0$ Mpc$^{-1}$.
\end{abstract}
\maketitle

\section{Introduction}
There are observational evidences which indicate that magnetic fields
exist not only in galaxies, but also in even larger systems, such as
cluster of galaxies and intra-cluster spaces \cite{2002RvMP...74..775W}.
Yet, the origin of such large scale magnetic fields is still a mystery
\cite{2002RvMP...74..775W}. It is now widely believed that the magnetic
fields at large scales are amplified from a tiny field and maintained by
the hydro-magnetic processes, i.e., the dynamo.  However, the dynamo
needs a seed field to act on and does not explain the origin of magnetic
fields itself. As far as the magnetic fields in galaxies are
concerned, the seed fields as large as $10^{-20} \sim 10^{-30}$ G are
required in order to account for the observed fields
\cite{1999PhRvD..60b1301D} of order $1 \mu$ G at the present universe.
Recent discovery of magnetic fields in galaxies at high redshifts
\cite{2008Natur.454..302B} may require even larger seed fields.

A classical way to generate seed magnetic fields in astrophysics is
based on the Biermann battery effect \cite{Biermann50}. The Biermann
battery works in various astrophysical systems, such as stars
\cite{1982PASP...94..627K}, supernova remnants
\cite{1998MNRAS.301..547M,Hanayama05}, protogalaxies
\cite{2000ApJ...540..755D}, large-scale structure formation
\cite{1997ApJ...480..481K}, and ionization fronts at cosmological
recombination \cite{2000ApJ...539..505G,1994MNRAS.271L..15S}.  These
studies show that magnetic fields with amplitude $10^{-16}$ $\sim$
$10^{-21}$ G could be generated.  However, the coherence-length of seed
fields generated by such astrophysical mechanisms may tend to be too
small to account for galaxy-scale magnetic fields.

On the other hand, cosmological mechanisms at the early inflationary
epoch can produce magnetic fields with a large coherence length since
accelerating expansion during inflation can stretch small-scale fields
to scales that can exceed the causal horizon
\cite{Ratra:1991bn,Bamba:2003av,2004PhRvD..70d3004P,Turner:1987bw}.
However, because no magnetic field is generated in simplest models with
the usual electromagnetic field, it is necessary to introduce some
extensions to the standard particle model.  Furthermore, it is recently
argued that the backreactions from the electro-magnetic
fields will stop the inflation
and significantly suppress the magnetic field generation
\cite{2009JCAP...08..025Dh,2009JCAP...12..009K}, if they  are
properly taken into account.

Generation mechanisms of magnetic fields in a decelerating universe
prior to cosmological recombination have also been proposed.
Originally, Harrison found that the vorticity in a primordial plasma can
generate magnetic fields \cite{1970MNRAS.147..279H}. This is because
electrons and ions would tend to spin at different rates as the universe
expands due to the radiation drag on electrons, arising a rotation-type
electric current and thus inducing magnetic fields.  Following his idea,
many authors have investigated magnetic field generation through the
second order vorticity generated from the first order density
perturbations
\cite{2005PhRvD..71d3502M,2004APh....21...59B,2005MNRAS.363..521G,2005PhRvL..95l1301T,2006Sci...311..827I,2010arXiv1012.2958F,2007astro.ph..1329I,2009CQGra..26m5014M,2006ApJ...651..627S}.

In this paper, we consider a first order vector (vorticity) mode firstly
investigated by Rebhan followed by Lewis
\cite{1992ApJ...392..385R,2004PhRvD..70d3518L}.  In a Friedmann universe
with a perfect fluid without anisotropic stress, the vector mode
has only a decaying mode. {This means a diverging vector metric
perturbation (frame-dragging potential) at the initial time and
therefore the model is inconsistent with an almost isotropic Friedmann
universe, and goes beyond the linear perturbation theory.}
However, in the existence of anisotropic
stress by free streaming particles such as neutrinos, it has been found
that there exists a regular (growing) mode with an initial non-zero
vorticity {and with isotropic initial phase space distributions}. We first present an observational constraint on this
primordial vector mode amplitude using the seven-year WMAP data. We then
estimate the amplitude of the magnetic fields inevitably created from this
vector mode in the light of the obtained constraint on the vector mode
amplitude.

This paper is organized as follows.
In Sec. II, the overview of the primordial vector mode investigated by
\cite{1992ApJ...392..385R,2004PhRvD..70d3518L} is described and we give
an analytic solution obtained by a tight coupling approximation.
In Sec. III, we show the constraints on the vector mode amplitude and
its spectral index from the seven-year WMAP data and discuss its implication. 
In Sec. IV, we calculate the magnetic field spectrum inevitably generated
from the primordial vector mode.
Finally, Sec. V is devoted to the summary and discussion. 

\section{Equations and Solutions}
\subsection{preliminary}
Here we quickly review basic equations for the evolution of primordial
vector modes.  We consider linear perturbations in the synchronous
gauge, in a flat Friedmann-Robertson-Walker universe with a metric
\begin{equation}
ds^2 = a(\eta)^2\left[-d\eta^2+(\delta_{ij}+h_{ij})dx^idx^j\right]~,
\end{equation}
where $a(\eta)$ is a scale factor. Because we are interested in the
vector mode, the metric perturbations $h_{ij}$ are decomposed in Fourier
space as
\begin{equation}
h_{ij}=i\hat{k}_i h^V_j+i\hat{k}_jh^V_i~,
\end{equation}
where $h_i^V$ is divergence-less, i.e., $\hat{k}^i h_i^V=0$. It is useful
to expand the vector $h^V_i$ in terms of two independent transverse
basis vectors as
\begin{equation}
h_i^V(\vec{k}) = \sum_{\pm} h^V_{\pm}(\vec{k}) e^{\pm}_{i}(\vec{k})~,
\end{equation}
where $\pm$ represents the parity. Because each parity component and
each Fourier mode evolve independently, we omit $\vec{k}$ and $\pm$
dependencies for simplicity in the rest of the paper.  Let us work with
the gauge invariant metric variable, $\sigma$, which is defined using the
synchronous gauge variable $h^V$ as $\sigma \equiv \dot{h}^V/k$.  Then,
the linearized Einstein equations give
\begin{eqnarray}
 k^2 \sigma&=&-16\pi Ga^2 \left(\bar{\rho}+\bar{P}\right)v~,\\
 \dot{\sigma}+2{\cal H} \sigma&=&8\pi Ga^2 \bar{P}\pi/k~ \label{eq:sigma},
\end{eqnarray}
where dot denotes a derivative with respect to the conformal time
$\eta$, ${\cal H}=\dot{a}/a$ is the conformal Hubble parameter, $\bar{\rho}$ and
$\bar{P}$ are the zero-th order density and pressure of the total fluid,
respectively, and $v$ and $\pi$ are the velocity and anisotropic
stress in the vector mode, respectively. From Eq.(\ref{eq:sigma}) it is
readily found that the vector metric perturbation has only a decaying mode
$\sigma \propto a^{-2}$ if there is no anisotropic stress.

For the matter part, the Euler equation of baryon is given by
\begin{equation}
\dot{v}_b+{\cal H}v_b=-\frac{4\rho_\gamma}{3\rho_b}an_e\sigma_T \left(v_b-v_\gamma\right)~,
\end{equation}
where $n_e$ is the electron number density and $\sigma_T$ is the Thomson
scattering cross section. For the photons we expand the distribution
function for the vector mode into multipoles and rewrite the Boltzmann 
equation as
\begin{eqnarray}
\dot{v}_\gamma +\frac{1}{8}k\pi_\gamma &=&
 -an_e\sigma_T\left(v_\gamma-v_b\right) ~,\\
\dot{\pi}_\gamma
+\frac{8}{5}kI_3-\frac{8}{5}kv_\gamma&=&-an_e\sigma_T\left(\frac{9}{10}\pi_\gamma-\frac{9}{5}E_2\right)-\frac{8}{5}k\sigma~,\\
\dot{I}_\ell+k\frac{\ell}{2\ell+1}\left(\frac{\ell+2}{\ell+1}I_{\ell+1}-I_{\ell-1}\right)&=&-an_e\sigma_T
 I_\ell ~~(\ell>2) ~,
\end{eqnarray}
where $I_\ell$ is the $\ell$-th order angular moment of the photon
distribution, and $E_\ell$ is the $\ell$-th order moment of the E-mode
polarization \cite{2004PhRvD..70d3518L}. 

\subsection{tight coupling approximation and solutions}

In the very early universe photons and baryons are tightly coupled
as the opacity $\frac{1}{\dot{\tau}}=an_e\sigma_T$ is large. This
enables us to expand the equations by a tight coupling parameter,
\begin{equation}
\epsilon=\frac{k}{\dot{\tau}}\sim 10^{-2}
\left(\frac{k}{1{\rm Mpc}}\right)
\left(\frac{1+z}{10^4}\right)^{-2}
\left(\frac{\Omega_b h^2}{0.02}\right)^{-1}~,
\label{eq:tcapara}
\end{equation}
where $\Omega_b$ is the baryon density normalized by the critical
density, and $h$ is the Hubble parameter normalized by $100$ km/s/Mpc.
At the lowest order, the equation for photon velocity is
\begin{equation}
\dot{v^{(0)}_\gamma}=-\frac{\dot{R}}{1+R}v^{(0)}_\gamma~,
\end{equation}
where $R=\frac{3\rho_b}{4\rho_\gamma}$. The solution is 
\begin{equation}
v^{(0)}_\gamma = \frac{v_{\gamma, \rm ini}}{1+R}~,
\label{eq:v0gamma}
\end{equation}
where $v_{\gamma, \rm ini}$ is the initial photon velocity. Therefore, the
fluid velocity stays constant deep in the radiation dominated era where
$R\ll 1$. At
this order the baryon-photon slip and the photon anisotropic stress do
not exist. At next order they are given by
\begin{eqnarray}
v^{(1)}_\gamma-v^{(1)}_b &=& \frac{k}{\dot{\tau}}\frac{R{\cal H}}{k(1+R)}v^{(0)}_\gamma~,\label{eq:(v-v)(1)}\\
\pi^{(1)}_\gamma &=& \frac{k}{\dot{\tau}}\frac{32}{15}\left(v^{(0)}_\gamma+\sigma^{(0)}\right)~,
\end{eqnarray}
respectively, where $\sigma^{(0)}$ is the solution of the vector metric perturbation
at the lowest order.  Note that the transfer of baryon-photon slip is
independent of $k$. In other words, because $\dot{R}\approx
R/\tau$ the slip term in the vector mode is further suppressed by a factor of
$R/(k\tau)$ on small scales. This is compared with the slip term
in the scalar mode, i.e. $v_\gamma-v_b \propto (k/\dot{\tau})R$, where the $k$ dependence
derives from the pressure gradient.

For the later discussion about magnetic field generation we derive the baryon-photon slip at the next order
in the tight coupling approximation. We find it to be
\begin{eqnarray}
v^{(2)}_\gamma-v^{(2)}_b 
&=&\frac{k}{\dot{\tau}}\frac{R{\cal H}}{(1+R)k}v^{(1)}_\gamma
-\frac{4}{15}
 \left(\frac{k}{\dot{\tau}}\right)^2
 \frac{R}{1+R}\left(v^{(0)}_\gamma
 +\sigma^{(0)}\right)\nonumber \\
&&+\frac{R^2}{(1+R)^2}
\frac{\cal H}{\dot{\tau}}\frac{k}{\dot{\tau}}\frac{v_\gamma^{(0)}}{k}
\left[\frac{\dot{R}}{R(1+R)}+\frac{\dot{\cal H}}{\cal H}
+\frac{\dot{v}_\gamma^{(0)}}{v_\gamma^{(0)}}
-\frac{\ddot{\tau}}{\dot{\tau}}+{\cal H}\right]~.
\label{eq:TCA2_part}
\end{eqnarray}
Therefore, up to the second order in the tight coupling approximation 
the baryon-photon slip is given by 
\begin{equation}
v_\gamma-v_b =
 \frac{k}{\dot{\tau}}\frac{R{\cal H}}{(1+R)k}v_\gamma-\frac{4}{15}
 \left(\frac{k}{\dot{\tau}}\right)^2\frac{R}{1+R}\left(v_\gamma+\sigma\right)~,
\label{eq:TCA2}
\end{equation}
where we have neglected cosmological redshift terms.
Note that the signs of the two terms are different. We shall see below
that, because of this,  a significant amount of magnetic fields resolves away
when the two terms become comparable around the Silk damping epoch.

\begin{figure}
\begin{minipage}[m]{0.45\textwidth}
\rotatebox{0}{\includegraphics[width=1.0\textwidth]{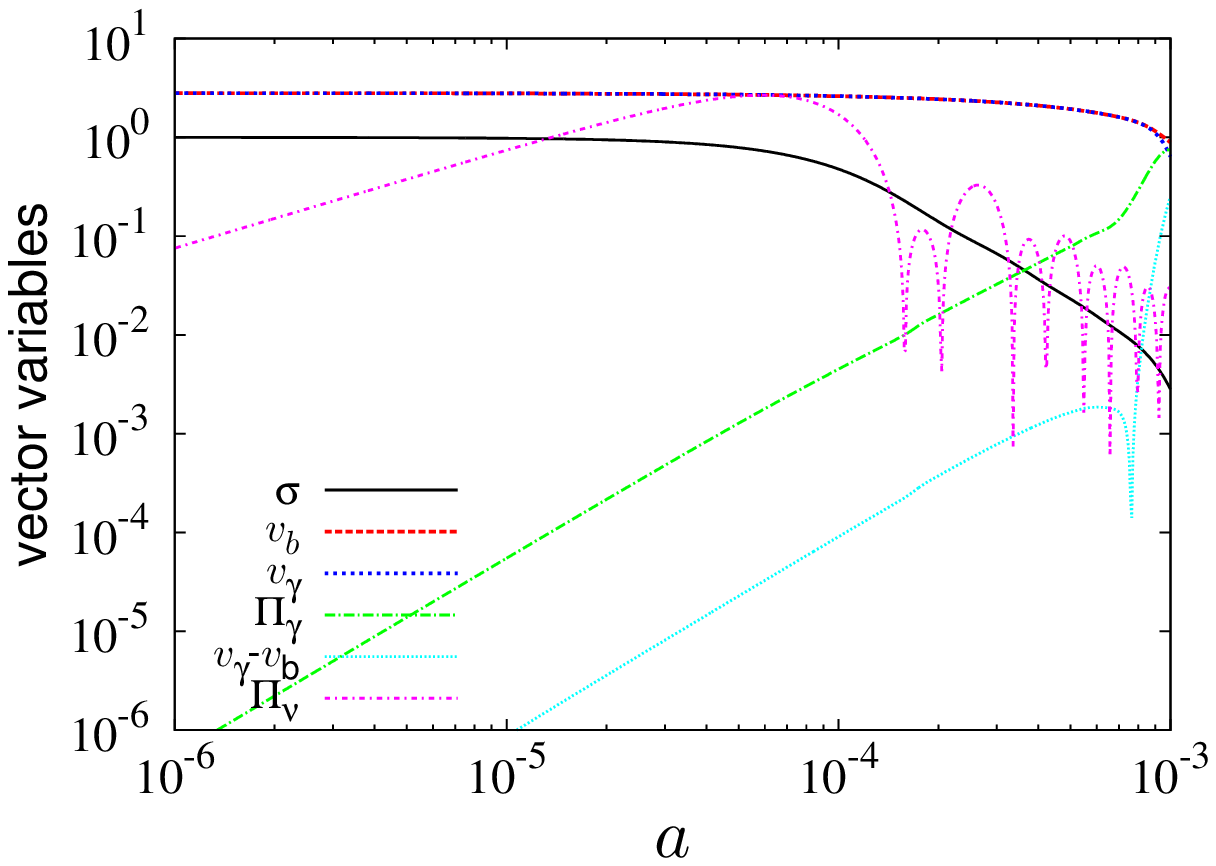}}
\end{minipage}
\begin{minipage}[m]{0.45\textwidth}
\rotatebox{0}{\includegraphics[width=1.0\textwidth]{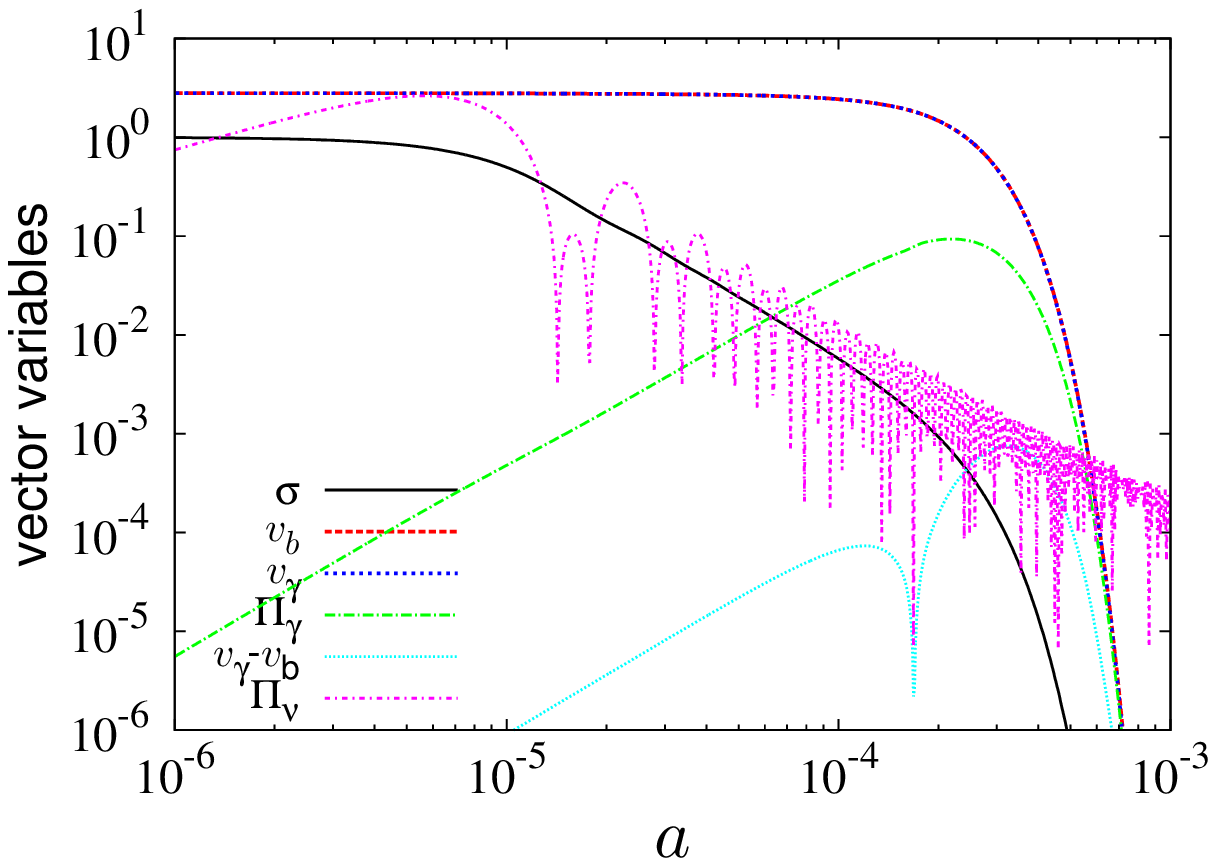}}
\end{minipage}
\caption{Time evolutions of the vector perturbation variables with
 wavenumber $k=0.1$ Mpc$^{-1}$ (left) and $k=1.0$ Mpc$^{-1}$ (right).
The vector potential $\sigma$ starts to decay after horizon crossing,
 while velocities $v_{b,\gamma}$ stay constant until the Silk damping takes place.}
\label{fig:fig1}
\end{figure}

{
Before moving to the next section we summarize the initial
conditions \cite{1992ApJ...392..385R,2004PhRvD..70d3518L}. By expanding
the equations in powers of $k\tau$ and assuming the radiation
dominated era, we find at the lowest order
}
\begin{eqnarray}
\sigma_{\rm ini}&=&\sigma_0~,\\
v_{\gamma, \rm ini} &=& \sigma_0\left(\frac{4R_\nu+5}{4R_\gamma}\right)~,\\
v_{\nu, \rm ini}&=&-\frac{R_\gamma}{R_\nu}v_{\gamma, \rm ini}~,\\
v_{b, \rm ini}&=&v_{\gamma, \rm ini}~,\\
\pi_{\nu, \rm ini}&=&-\frac{2}{R_\nu}\sigma_0 k\tau~,
\end{eqnarray}
where we have defined the ratios
$R_\nu=\frac{\rho_\nu}{\rho_\nu+\rho_\gamma}$ and
$R_\gamma=1-R_\nu$. Evolutions of the vector perturbations are presented
in Fig. \ref{fig:fig1}.

\section{CMB Constraint on the primordial vector mode}
In order to compare with the CMB data precisely, we use the CosmoMC
package with a modification to include two new parameters, $r_v$ and
$n_v$. Here $r_v$ is the vector-scalar ratio and $n_v$ is the spectral
index of the vector power spectrum. Specifically, we parameterize the
power spectrum of primordial vector mode with an initial amplitude of
the metric perturbation $\sigma$ as
\begin{equation}
\frac{k^3}{2\pi^2}\left<\sigma(\vec{k})\sigma^\ast(\vec{k}^\prime)
\right>={\cal P}_\sigma(k)\delta(\vec{k}-\vec{k}^\prime)=A_\sigma\left(\frac{k}{k_0}\right)^{n_v-1}\delta(\vec{k}-\vec{k}^\prime)~,
\end{equation}
where $k_0=0.002$ Mpc$^{-1}$ is the pivot scale. The vector-scalar ratio
is defined by
\begin{equation}
r_v = \frac{A_\sigma}{A_s}~,
\end{equation}
where $A_s$ is the scalar counterpart of the power spectrum amplitude.

The likelihood function we calculate is given by the WMAP collaboration
\cite{2011ApJS..192...14J,2011ApJS..192...16L} and the codes are
publicly available at their web site.  To include
the primordial vector mode, we calculate the angular power spectrum,
$C_{\rm vector ~ \ell}$, of the CMB anisotropy of temperature
and E-mode polarization auto-correlations (TT and EE) and their cross
correlation (TE) using the CAMB code \cite{Lewis:1999bs}. We add these power 
spectra to those from the scalar and tensor modes as
\begin{equation}
C_{\rm tot~ \ell}^{\rm TT,TE,EE}=C_{\rm scalar~\ell}^{\rm TT,TE,EE}+C_{\rm tensor~\ell}^{\rm TT,TE,EE}+C_{\rm vector~\ell}^{\rm TT,TE,EE}~,
\end{equation}
where we have assumed that the primordial scalar, vector and tensor
modes are statistically independent.  Then $C_{\rm tot~ \ell}$ are
fitted to the CMB data. The B-mode polarization angular power spectrum
is not used for our analysis because the sensitivity of the current
observational data of B-mode polarization is not enough to give limits
on the vector or tensor mode amplitudes.

\begin{figure}
\begin{minipage}[m]{0.45\textwidth}
\rotatebox{0}{\includegraphics[width=1.0\textwidth]{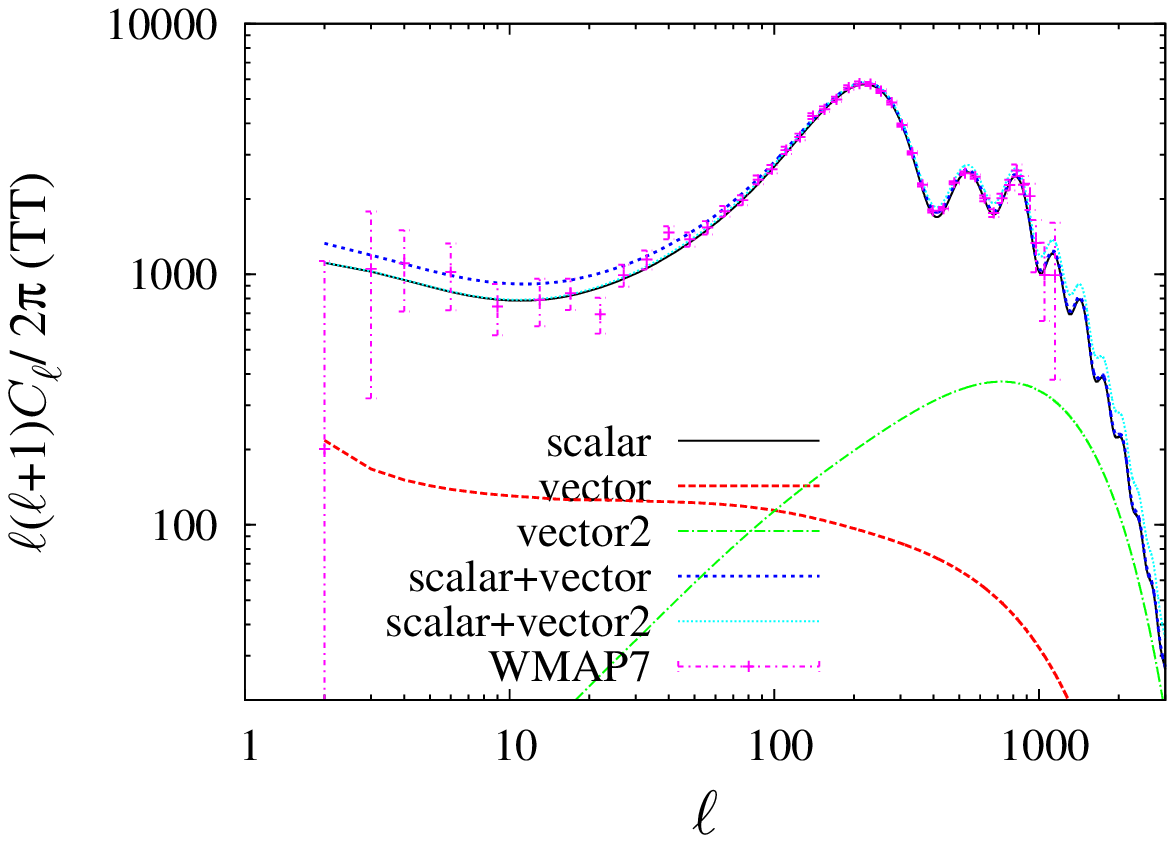}}
\end{minipage}
\begin{minipage}[m]{0.45\textwidth}
\rotatebox{0}{\includegraphics[width=1.0\textwidth]{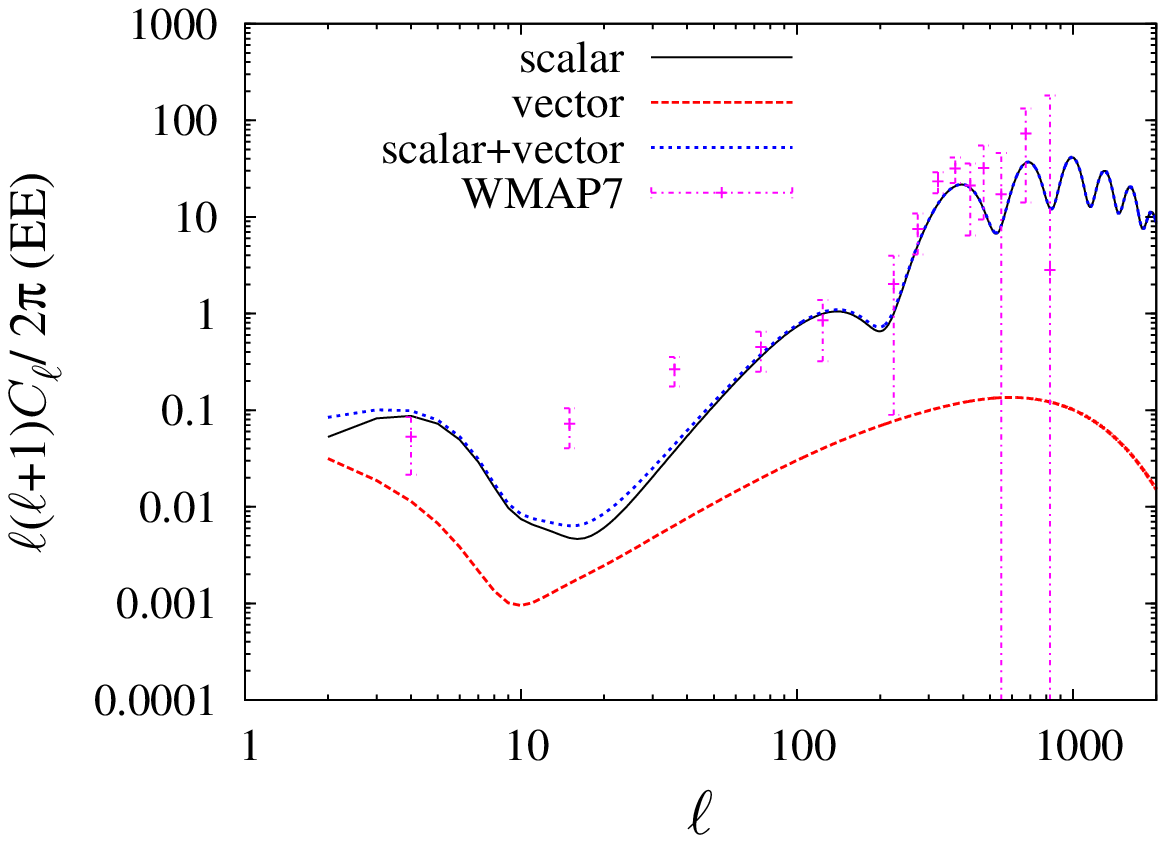}}
\end{minipage}
\begin{minipage}[m]{0.45\textwidth}
\rotatebox{0}{\includegraphics[width=1.0\textwidth]{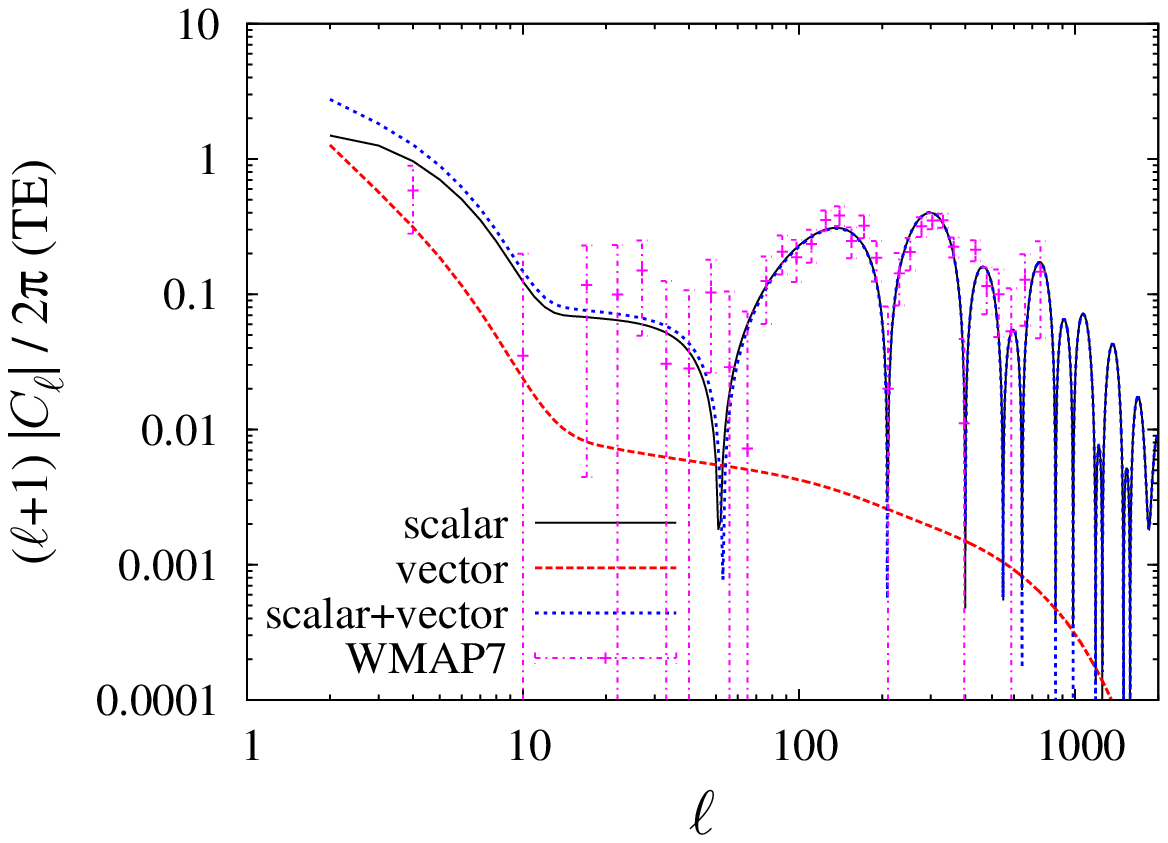}}
\end{minipage}
\caption{CMB angular anisotropies power spectra, temperature (TT),
 E-mode polarization (EE), and their cross-correlation (TE), induced by
 scalar and vector type perturbations. The constraint mainly comes from
 the TT angular power spectrum. The vector mode parameters are taken as
 $(r_v,n_v)=(0.945\times 10^{-2},0.921)$ and $(1.0\times 10^{-3},2.0)$
 for red dashed lines and green dash-dotted line (TT spectrum only), respectively.}
\label{fig:Cls}
\end{figure}

In Fig \ref{fig:Cls}, we depict the CMB angular power spectra along with
the seven-year WMAP data. In that figure, the parameters of the
primordial vector mode are fixed to $r_v=9.45\times 10^{-3}$ and
$n_v=0.921$, respectively, which are the allowed values at $95$\%
confidence levels (red dashed lines in Fig.\ref{fig:Cls}). The spectra
look similar to those from the tensor mode perturbations, although the
powers extend to the higher multipoles without oscillatory features.  As
clearly seen from the figure, the current constraints are mainly placed
on from the TT power spectrum at low multipoles if the vector mode is
nearly scale invariant. The situation is the same with the case of the
current constraint on the primordial tensor mode (gravitational waves)
from the CMB power spectrum. Only in the panel for the TT power spectrum
we also show the case with a bluer spectral index $n_v=2.0$ with
$r_v=1.0\times 10^{-3}$, which are also at the edge of $95$\% confidence
levels. In this case, we find that the constraint again comes from TT
power spectrum, but at the higher multipoles around $\ell \approx 1000$.

\begin{figure}
\begin{minipage}[m]{0.48\textwidth}
\rotatebox{0}{\includegraphics[width=0.99\textwidth]{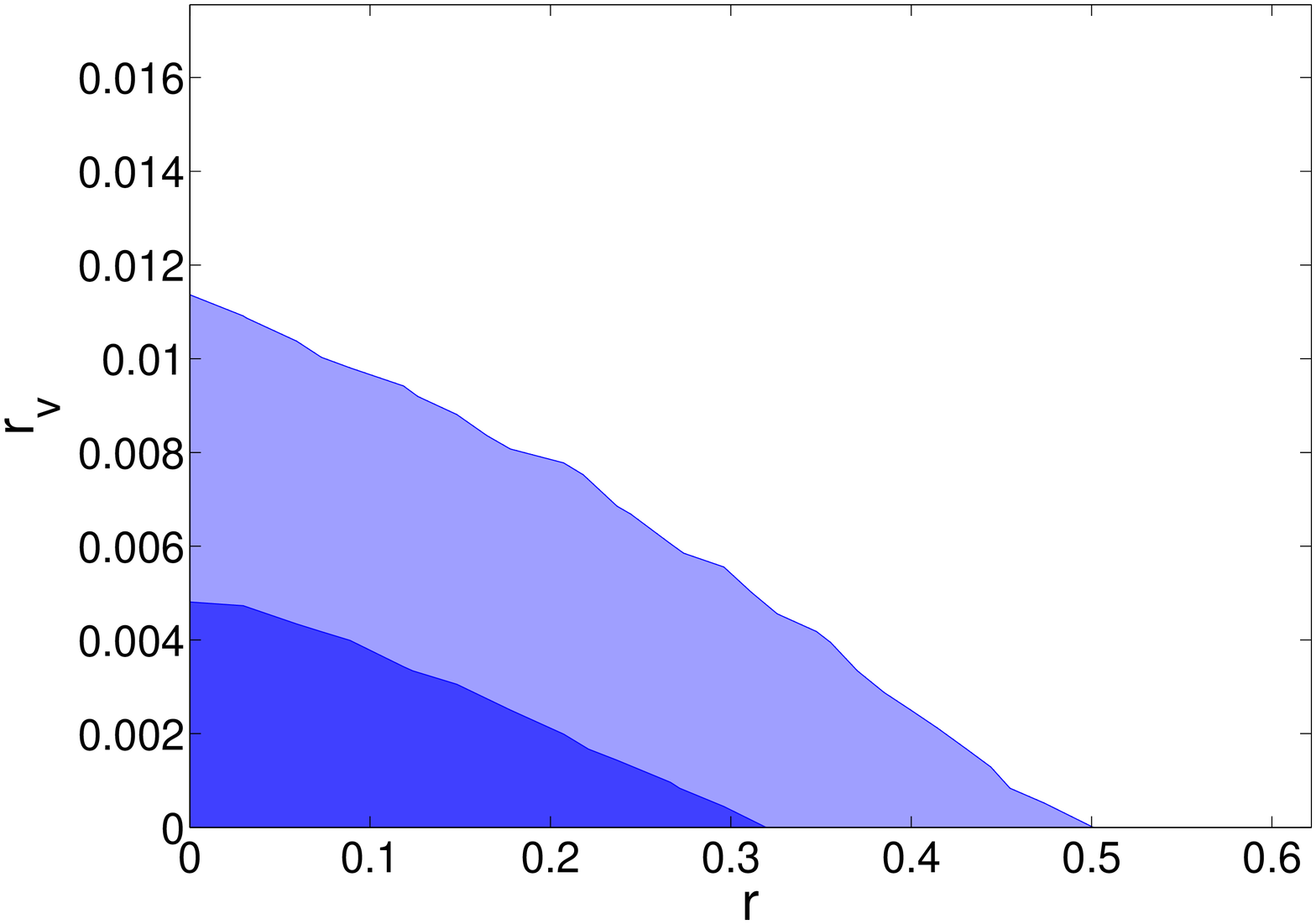}}
\end{minipage}
\begin{minipage}[m]{0.489\textwidth}
\rotatebox{0}{\includegraphics[width=0.99\textwidth]{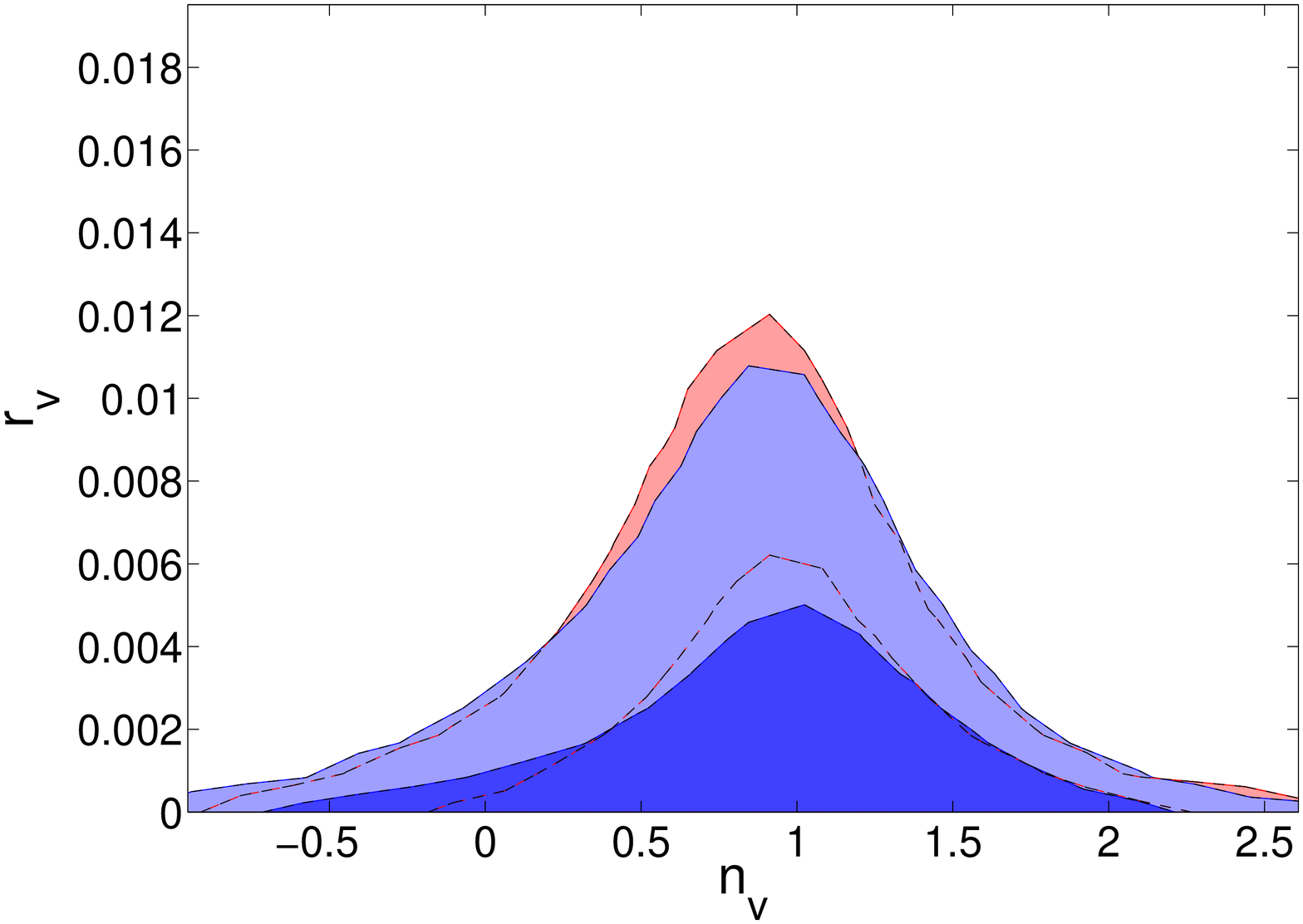}}
\end{minipage}
\caption{68 \% and 95 \% confidence regions from WMAP 7-yr data on
 $r$-$r_v$ (left) and $n_v$-$r_v$ planes (right). Left panel shows that
 constraint is on the total amount of vector and scalar
 components. Right panel shows constraints become severer when vector
 perturbation spectrum gets away from scale invariant $n_v=1$. In that
 panel red contours are obtained from the analysis without tensor modes, 
 blue contours are obtained including tensors. Blue and
 red spectra are allowed when $r_v\to 0$.}
\label{fig:cont}
\end{figure}

The correlation between the parameters of primordial vector modes, $r_v$
and $n_v$, is shown in the right panel of Fig \ref{fig:cont}. If the
primordial vector perturbation is given almost scale invariant, we found
that the constraint is put at lowest multipoles as mentioned above and
the constraint is the weakest. When the spectrum becomes bluer
as $n_v\gtrsim 1$, the constraint comes from the higher multipoles
and hence it becomes tighter. For example,
when $n_v\approx 2.0$ the constraint becomes
$r_v\lesssim 0.001$, while $r_v\lesssim 0.009$ when $n_v\approx 1.0$.

To make our analysis as general as possible, we made our Markov chain
analysis with and without the primordial
tensor mode. The effect of including the tensor mode on the constraint
on the vector mode is also shown in the right panel of
Fig. \ref{fig:cont}. We observe that the constraint on the vector mode
generally becomes tighter if the tensor mode is included. The 
constraints we found are $r_v\lesssim 9.45\times 10^{-3}$ (without
tensor) and $r_v\lesssim 8.36\times 10^{-3}$ (with tensor) when the all
the other parameters are marginalized.
The simultaneous constraint on the vector and tensor modes is shown in
the left panel of Fig. \ref{fig:cont}. The result shown is easily
understood: because the angular power spectra of temperature anisotropies
from the vector and tensor modes are similar, the current CMB data can
put constraint on the total amount of the vector and tensor
mode perturbations. We found the constraint to be
\begin{equation}
r_v + \left(\frac{r}{40}\right)\lesssim 0.012 ~,
\end{equation}
at $95.4$\% confidence level where $r$ stands for the tensor-scalar ratio.

\section{magnetic field generation}
When there exists velocity difference between baryons and photons in the
vector mode, magnetic fields may be generated inevitably. This arises
because photons scatter off electrons preferable to protons. Electric
fields are induced to prevent charge separation between electrons and
protons, and magnetic fields are generated from these induced electric
fields through Maxwell equations if there exists the vorticity
difference.  In general, magnetic fields then affect the evolution of
the fluids. However, we omit any backreactions from magnetic fields on
the fluid motion because we are interested in the magnetic field
generation from zero and thus the backreactions from the magnetic fields
will be negligible. In terms of the cosmological perturbation theory,
the backreactions from magnetic fields are second order, namely, the
Lorentz force would be $vB\sim {\cal O}(v^2)$ and the energy momentum
tensor will be proportional to $B^2 \sim {\cal O}(v^2)$.

The evolution equation of magnetic fields due to the Thomson scattering
is given by
\begin{equation}
\frac{d (a^2 B^i)}{dt} = \frac{4\sigma_T \rho_\gamma a}{3e}\epsilon^{ijk}
 \left(v_{\gamma~ j,k} -v_{b~ j,k}\right)~,
\label{eq:dBdt}
\end{equation}
where $\epsilon^{ijk}$ is the Levi-Civita tensor and $e$ is the electric
charge. For the scalar type perturbation, this term vanishes because
$\epsilon^{ijk}v^S_{j,k}\sim \epsilon^{ijk}\hat{k}_j\hat{k}_k v^S=0$. In
this paper we solve the above equation with the initial condition
$B=0$ at $z=10^{9}$, which roughly corresponds the time of neutrino decoupling.

\begin{figure}
\begin{minipage}[m]{0.49\textwidth}
\rotatebox{0}{\includegraphics[width=1.0\textwidth]{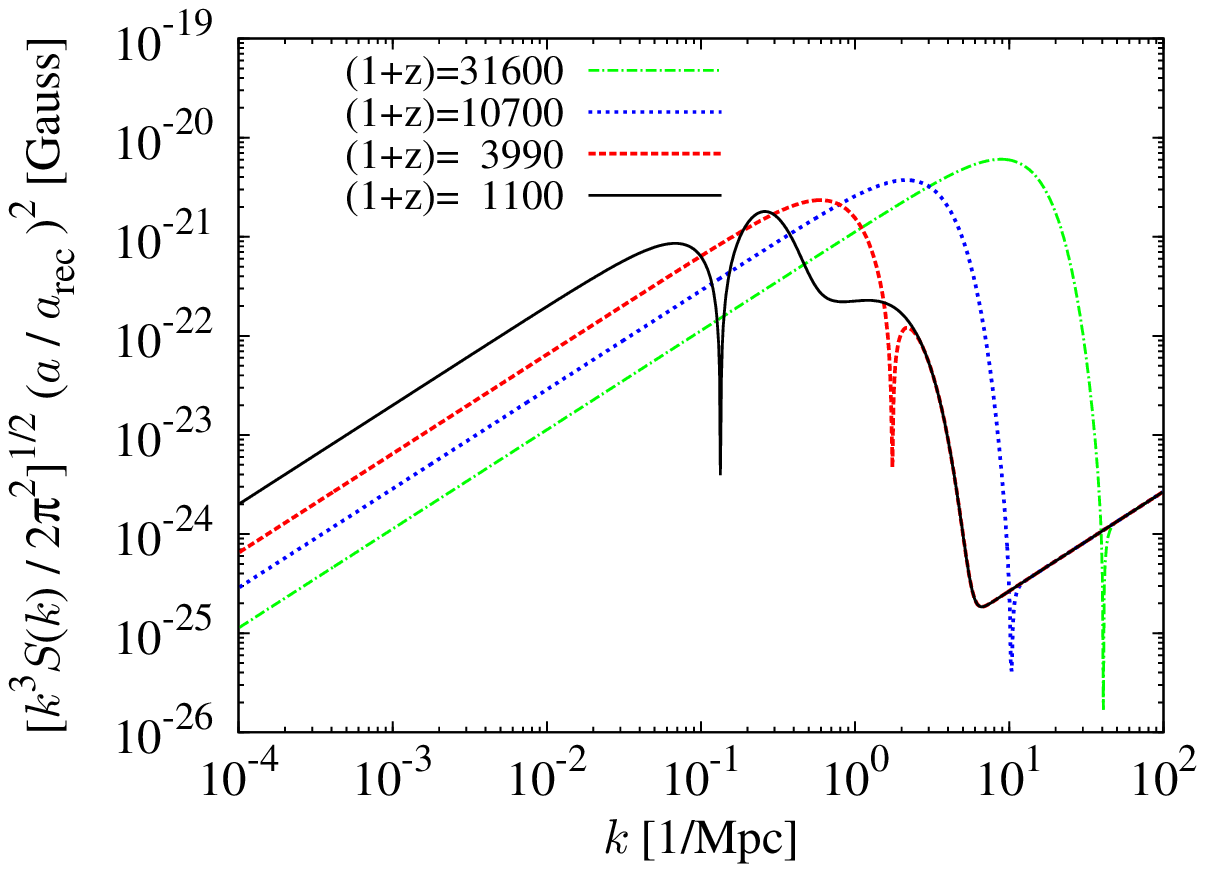}}
\end{minipage}
\begin{minipage}[m]{0.49\textwidth}
\rotatebox{0}{\includegraphics[width=1.0\textwidth]{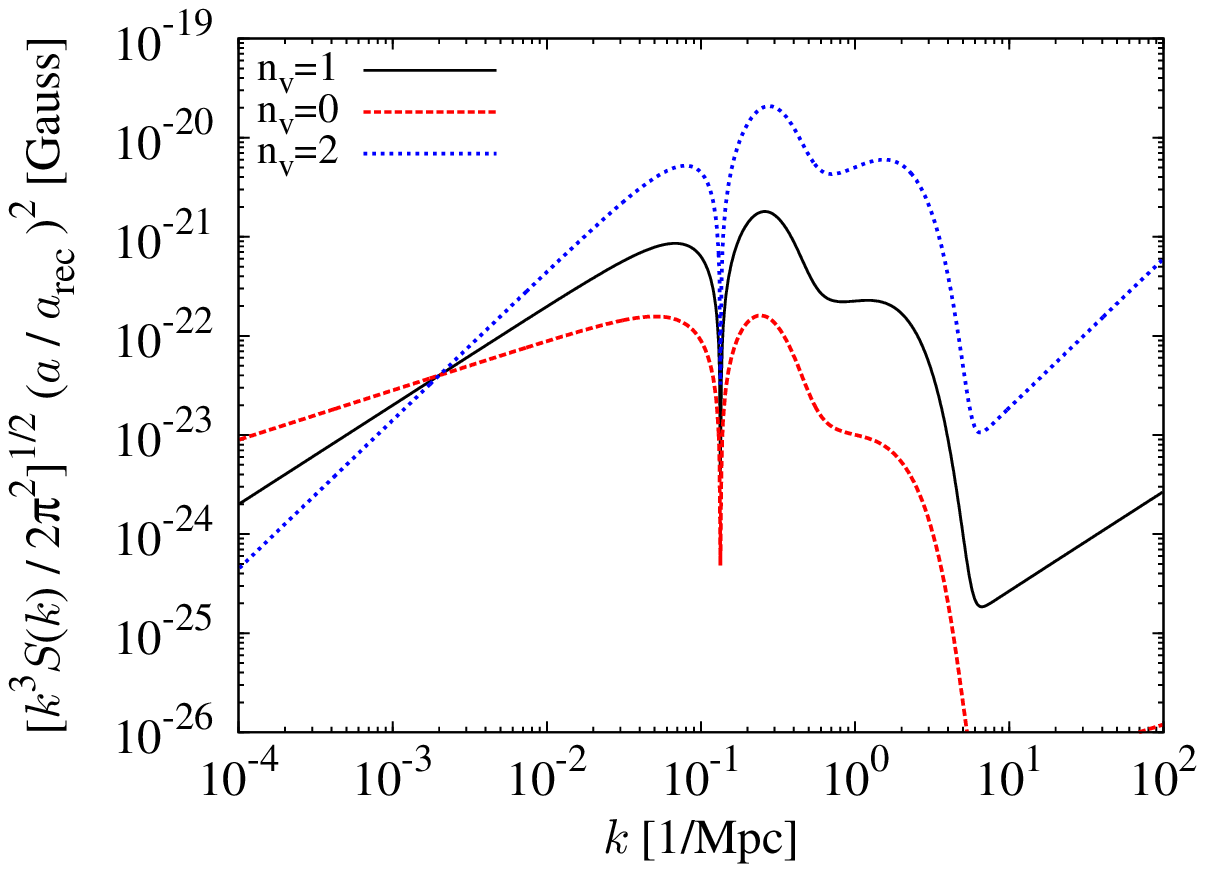}}
\end{minipage}
\caption{Left: The spectra of magnetic fields generated from primordial vector
 modes at several times as indicated in the figure. The final
 spectrum at decoupling epoch is shown by the black solid line.
Right: The dependence on $n_v$ at $z=1100$.}
\label{fig:mag}
\end{figure}

Even in the case where the vector mode perturbation exists, the mean of
the magnetic fields should still be zero, but there exists the variance.
To find the spectrum of magnetic fields, we first calculate $B^i B_i$.
That is given by 
\begin{equation}
a^4 B^i(\vec{k},t)B^\ast_i(\vec{k^\prime},t)
=\left(\frac{4\sigma_T}{3e}\right)^2
\left(\delta^{j\ell}\delta^{km}-\delta^{jm}\delta^{k\ell}\right)k_k k_m^\prime 
\int\int a(t^\prime)\delta v_j(\vec{k},t^\prime)
a(t^{\prime\prime})\delta v^\ast_\ell(\vec{k},t^{\prime\prime})dt^\prime dt^{\prime\prime}~.
\end{equation}
Taking the ensemble average of this expression and defining the vector
power spectrum as
\begin{eqnarray}
\left<
\delta v_j(\vec{k},t^\prime)\delta v^\ast_\ell(\vec{k^\prime},t^{\prime\prime})
\right>
&=&\frac{2\pi^2}{k^3}{\cal P}_\sigma(k)P_{j\ell}(\hat{k})\delta v(k,t^\prime)\delta
v(k,t^{\prime\prime})\delta(\vec{k}-\vec{k^\prime})\\
P_{j\ell}(\hat{k})&=&\delta_{j\ell}-\hat{k}_j\hat{k}_\ell~,
\end{eqnarray}
where $\delta v(k,t)$ is the transfer function of baryon-photon slip with 
$\sigma=1$ at the initial time, and the magnetic field power spectrum can be written as
\begin{equation}
a^4(t)\frac{k^3}{2\pi^2}S_B(k,t)=\left(\frac{4\sigma_T}{3e}\right)^2
 2{\cal P}_\sigma(k) k^2
\left[
\int dt^\prime a(t^\prime)\rho_\gamma(t^\prime)\delta v(k,t^\prime)
\right]^2~.
\label{eq:mag_field_spectrum}
\end{equation}
Here, magnetic field power spectrum $S_B(k)$ is defined as
\begin{equation}
S_B(k)\delta(\vec{k}-\vec{k}^\prime)
=\left<B^i(\vec{k})B_i(\vec{k}^\prime)\right>~.
\end{equation}
We numerically calculate the magnetic field spectra and show them in
Fig. \ref{fig:mag}. In that figure, we took the vector mode parameters
as $r_v=0.01$ and $n_v=1$, which give the maximum amount of the vector mode
allowed at large scales.

First of all, let us roughly estimate the amplitude of the generated
magnetic fields. 
From Eq.(\ref{eq:dBdt}), it is estimated as
\begin{eqnarray}
(a^2 B) &\simeq& \left(\frac{4\sigma_T}{3e}\right)\frac{a}{H}\rho_\gamma k
 (v_\gamma-v_b)~,\nonumber \\
&\simeq&\left(\frac{4\sigma_T}{3e}\right)a^2\rho_\gamma\left(\frac{k}{\dot\tau}\right)\frac{R}{(1+R)^2}\left(\frac{4R_\nu+5}{4R_\gamma}\right)\sigma_0,\nonumber \\
&\sim&\left(\frac{4\sigma_T}{3e}\right)a^2\rho_\gamma\left(\frac{k}{\dot\tau}\right)\sigma_0
\label{eq:32}
\end{eqnarray}
where $H$ is the usual Hubble parameter and we have used the tight
coupling solution for $\left(v_\gamma-v_b\right)$ derived in Sec. II. By substituting
Eq.(\ref{eq:tcapara}), $\rho_\gamma\simeq 2\times 10^{-51}(1+z)^4$
[GeV$^4$], and $\sigma_T\simeq 1.7\times 10^{3}$ [GeV$^{-2}$], we find
\begin{equation}
a^2B \sim 1.2\times 10^{-27} \mbox{G} 
\left(\frac{k}{{\rm Mpc}^{-1}}\right)
\left(\frac{1+z}{10^4}\right)^{-1}
\left(\frac{r_v}{0.01}\right)^{1/2}~.
\end{equation}
Therefore, around the cosmological recombination epoch, $B\sim 10^{-21}$ G
is expected for the magnetic field strength.

At super-horizon scales, the magnetic fields have a power-law
$B\propto k$ if the vector mode power spectrum is scale invariant.
This is manifestly shown by Eq.(\ref{eq:32}).
If the primordial vector mode is tilted
as $n_v\neq 1$,
the magnetic field spectrum at super horizon scale should become as
$B\propto k^{(n_v+1)/2}$.

At subhorizon scales, we found some interesting features in the
resultant magnetic field spectrum.  First, in the radiation dominated
era, there is a peak just above the Silk damping scale. Secondly, we
find a characteristic cut-off below the Silk damping scale. These
features were not observed in the magnetic field spectrum from the
second order vector modes. In that case the magnetic field spectrum is
extended toward much smaller scales
\cite{2007astro.ph..1329I,2009CQGra..26m5014M,2010arXiv1012.2958F}.  In
the primordial vector mode considered here there is no significant
difference in the evolution of velocity difference between photons and
baryons before and after the horizon crossing. Therefore the magnetic
fields continue to grow after the horizon crossing, which is manifestly
shown in the magnetic field spectrum as $B\propto k$ for $k\gtrsim k_h$,
with $k_h$ being the wavenumber corresponding to the Hubble scale. It is
expected that the generation of magnetic fields ceases around the Silk
damping epoch where the velocity difference, $v_\gamma - v_b$, reaches
its maximum value and starts to diminish. The peak position and the
amplitude of magnetic fields there can be estimated as follows.  The
magnetic fields generated from Eq.(\ref{eq:dBdt}) is estimated as
\begin{equation}
{\cal B}\equiv (a^2 B) \approx \frac{a}{H}\rho_\gamma k (v_\gamma-v_b)\propto ka~,
\end{equation}
where we have defined the comoving magnetic field ${\cal B}$, and used
the relations in the radiation dominated era such as $H\propto a^{-2}$,
$\rho_\gamma\propto a^{-4}$, and $(v_\gamma-v_b)\propto a^2$. The
comoving magnetic fields evolve in time as $\propto a$ and therefore the maximum
value is reached at the Silk scale where the perturbations start to be
erased. Because the diffusion scale is scaled as $k_{\rm diff}\propto
a^{-3/2}$ \cite{1995ApJ...444..489H}, the scale factor when diffusion
damping occurs at a given scale $k$ is given as $a_{\rm Silk}\propto
k^{-2/3}$. Therefore, if the magnetic field generation just ceased when
the Silk damping started we should expect that the magnetic fields have
the spectrum as ${\cal B}\propto k a_{\rm Silk}\propto k^{1/3}$. In
Fig. \ref{fig:mag} we find that the peak amplitudes at different
redshifts are indeed on this scaling relation.

However, we found a non-trivial cancellation of the magnetic field
generation. In fact, the magnetic field generation does not only cease
at the Silk damping epoch, but a significant amount of magnetic fields
generated by that time vanishes as is shown in the right panel of Fig. 5. This happens because the sign of the
velocity difference, $v_\gamma-v_b$, flips at the onset of the Silk
damping, which is shown in the left panel of Fig. 5. Deep in the
radiation dominated era, the velocity of radiation (photon) fluid stays
constant due to the redshift of energy density (Eq. (\ref{eq:v0gamma})),
while the velocity of baryon fluid should decay as $v_b \propto a^{-1}$
if baryons had no interaction with photons. Therefore during that time
the baryon fluid is dragged by the photon fluid, and we expect that
$v_\gamma-v_b>0$ if $v_{\gamma,{\rm ini}}>0$.  {After the diffusion
damping starts to erase the perturbations in photons, the baryon fluid,
which will try to keep rotating by its inertia, drags the photons so
that $v_\gamma-v_b<0$ if $v_{\gamma,{\rm ini}}>0$.  } This argument is
consistent with the result obtained from the tight coupling expansions,
in other words, the magnetic fields generated by the first term in
Eq. (\ref{eq:TCA2}) are largely compensated by the second term. The
resultant (comoving) magnetic fields spectrum has the characteristic
cut-off at $k\gtrsim 1$ [Mpc$^{-1}$].

\begin{figure}[ht]
\begin{minipage}[m]{0.49\textwidth}
\rotatebox{0}{\includegraphics[width=1.0\textwidth]{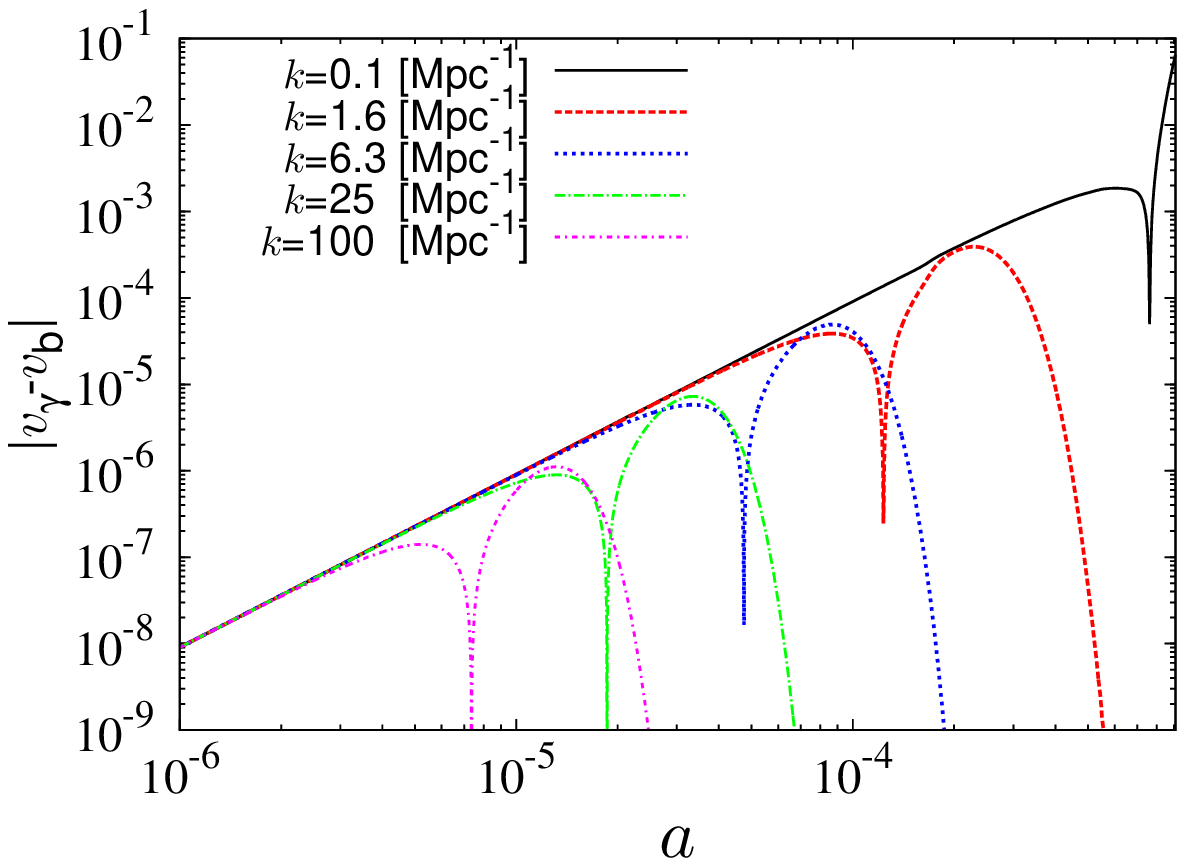}}
\end{minipage}
\begin{minipage}[m]{0.49\textwidth}
\rotatebox{0}{\includegraphics[width=1.0\textwidth]{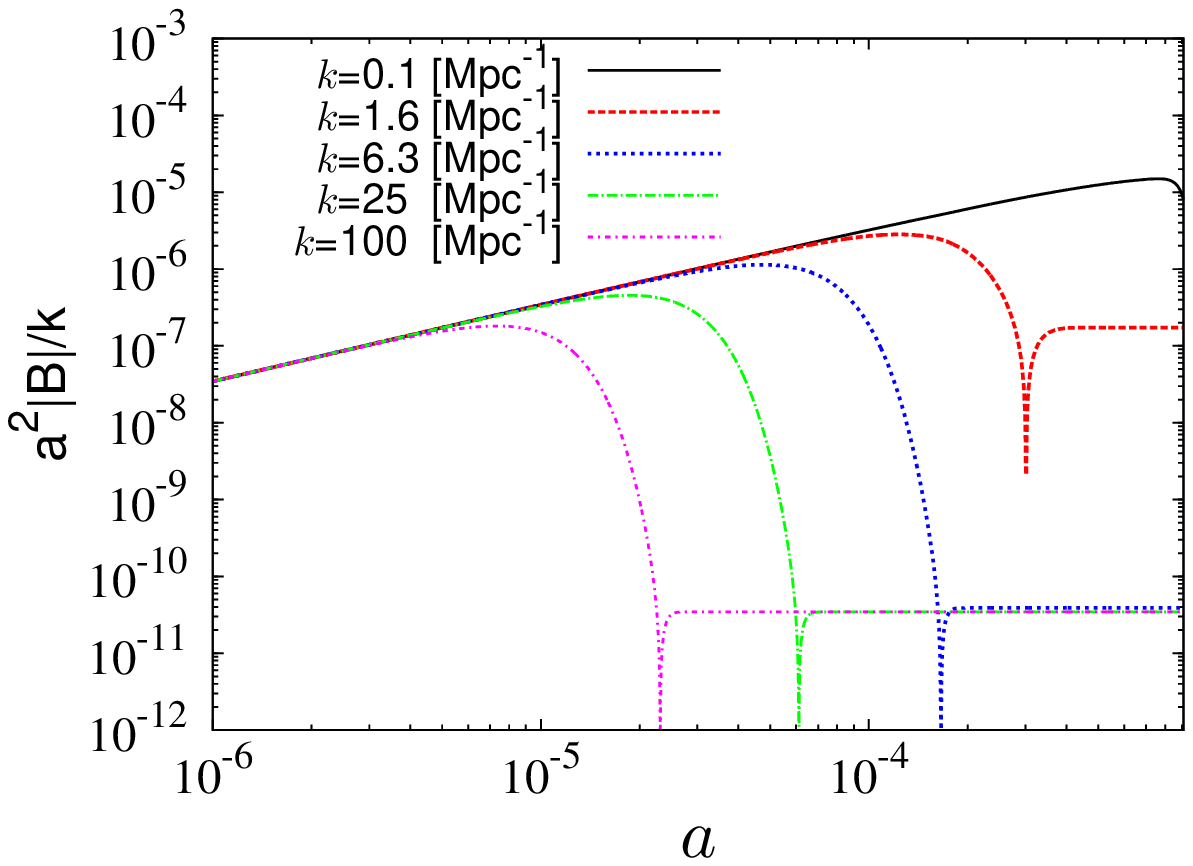}}
\end{minipage}
\caption{Time evolutions of the source for the magnetic field (left) and
 the transfer of the magnetic field (right) for some wavenumbers as
 indicated. At first $v_\gamma-v_b = \alpha v_\gamma$ with
 $\alpha>0$. This is because $v_\gamma \propto$ const. and $v_b\propto
 a^{-1}$ in the absence of interactions between photons and baryons, and
 therefore photons drag the baryons through Thomson interactions (the
 first term in Eq. (\ref{eq:TCA2})).}  \label{fig:fig2}
\end{figure}

\section{Summary and Discussion}
In this paper we examined observational constraints on the primordial
vector mode discussed in \cite{2004PhRvD..70d3518L}. This vector mode is
a vector analogue of neutrino isocurvature velocity scalar mode
\cite{2000PhRvD..62h3508B} and has a non-decaying solution of vector
metric perturbation. The vector mode perturbations generate CMB angular
anisotropies mainly through the Doppler effect, whose angular power
spectrum is similar to that from the tensor mode without oscillatory
features at small angular scales. Thus the amplitude of vector mode
perturbations is constrained from CMB data. We used seven-year WMAP data to
place a constraint on the vector-scalar ratio $r_v$. The constraint we
obtained is $r_v \lesssim 0.01$ when the vector spectrum is nearly scale
invariant. If the tensor mode is included the constraint is generalized
to $r_v+(r/40) \lesssim 0.012$ where $r$ is tensor-scalar ratio. The
current constraint is placed mainly from the temperature anisotropy
data. Future precise polarization data will tighten this bound
significantly, down to $r_v \lesssim 10^{-3}$ for Planck
\cite{2004PhRvD..70d3518L}.

As discussed by several authors, magnetic fields are generated
before cosmological recombination through Compton scatterings if there
existed cosmological vector mode perturbations. We numerically calculate
the magnetic field spectrum generated from such primordial vector
mode. We found that the spectrum is $B \approx 10^{-23}{\rm G}
\left(\frac{r_v}{0.01}\right)^{1/2}\left(\frac{k}{0.002}\right)^{(n_v+1)/2}$
at cosmological recombination where $n_v$ is the spectral index of the
primordial vector mode.  The magnetic fields monotonically increases
with wavenumber as $B\propto k^{(n_v+1)/2}$ up to the Silk damping
scale. Below this scale we found that there is a non-trivial
cancellation in the magnetic field generation, leading around five
orders of magnitude decrease in the magnetic field amplitude. The cancellation
occurs because the baryons drag the photons around the Silk damping
epoch, before which, on the contrary, the photons dragged the
baryons.

{In fact, perhaps surprisingly, using the second order tight coupling approximation we
can show that the magnetic fields should eventually vanish at this order
as follows. Deep in the radiation dominated era and at sub-horizon
scales, the tight coupling equation of photon velocity is written as
\begin{equation}
v_\gamma^\prime + \frac{k^2}{\dot\tau}\frac{4}{15}v_\gamma=0~.
\end{equation}
This equation can be solved to give the solution
\begin{equation}
v_\gamma = v_{\gamma, {\rm ini}} \exp\left(-\frac{4k^2}{45\beta}\eta^3\right)~,
\end{equation}
where we have introduced a constant $\beta$ to describe the differential
optical depth as $\dot\tau \equiv \beta \eta^{-2}$ in the radiation
dominated era. Inserting this
solution into the tight coupling expression of the baryon-photon slip
(Eq.(\ref{eq:TCA2})), and writing the time dependences of $a$, $R$ and
$\rho_\gamma$ as $a=\alpha\eta$, $R=R_0\eta$ and $\rho_\gamma=\gamma\eta^{-4}$,
respectively, the time integral to get the magnetic field amplitude can
be expressed as
\begin{equation}
(a^2B)\simeq \left(\frac{4\sigma_Tk}{3e}\right)\int_{\eta_{\rm ini}}^{\eta_{\rm end}} a^2 \rho_\gamma
 \delta v d\eta = \left(\frac{4\sigma_Tk}{3e}\right)\int_{\eta_{\rm ini}}^{\eta_{\rm end}} \alpha^2
 \left(\frac{\gamma}{\beta}\right)R_0 v_{\gamma, {\rm ini}}
\left[
1-\frac{4}{15}{k^2}{\beta}\eta^3 \exp\left(-\frac{4k^2}{45\beta}\eta^3\right)
\right]d\eta~.
\end{equation}
}
{The above integral of the form $\int_{\eta_{\rm ini}}^{\eta_{\rm end}
}(1-K\eta^3)\exp(-K\eta^3/3)d\eta=\left[\eta\exp\left(-\frac{1}{3}K\eta^3\right)\right]_{\eta_{\rm
ini}}^{\eta_{\rm end}}$ (with K constant)
goes exactly to zero for $\eta_{\rm ini} \to 0$ and $\eta_{\rm end} \to
\infty$. }

{The cancellation is perfect only within the second order tight
coupling approximation in the completely radiation dominated era, and only when we integrate from $\eta=0$ to $\eta=\infty$. Because we assume
that the magnetic field generation starts at some finite time,
$B(\eta_{\rm ini})=0$ with $\eta_{\rm ini}= \eta(z=10^{9})$,
the surface term should remain as
\begin{equation}
a^2B\simeq -\left(\frac{4\sigma_Tk}{3e}\right)\alpha^2 \gamma
 \left(\frac{R_0}{\beta}\right)v_{\gamma,{\rm ini}}\eta_{\rm
 ini}=-\left(\frac{4\sigma_Tk}{3e}\right) a^2 \rho_\gamma \frac{R}{\dot
 \tau}v_{\gamma,{\rm ini}}~.
\end{equation}
We have found that the surface term above dominates the magnetic fields
at small scales 
over the higher order terms in the tight coupling approximation, and
indeed, our numerical solution is consistent with the above estimate at $k\gtrsim 10$ Mpc$^{-1}$. This suggests that the magnetic fields start to increase
again with wavenumber as $B\propto k^{(n_v+1)/2}$ which is shown in
Fig. \ref{fig:mag}.
} Consequently, magnetic fields will have a small-scale
power up to $\sim 2\times 10^{-3}$ pc in the comoving scale. This scale
corresponds to the damping scale when the temperature of the universe
was around MeV, after which epoch neutrinos can free-stream and the
primordial vector mode solution considered here can apply.

The vector modes and magnetic fields considered in this paper are
different from the recently investigated second order vector modes or
magnetic fields
\cite{2008PhRvD..77d3523L,2009JCAP...02..023L,2010arXiv1008.4866C}. In
those studies the authors considered the second order
vector modes and/or magnetic fields generated from the non-linear
couplings of the first order scalar (density) modes. In the framework of the
cosmological perturbation theory, the second
order solution may be considered as a particular solution (in the limit
of neglecting vector-scalar couplings), because the couplings of first order
density modes can be considered as a source term for the otherwise
homogeneous evolution equations of the vector mode. Therefore,
the general solution may be expressed as a superposition of the
solutions, namely, the second order solution and the solution considered
in this paper. The second order solution suggests that magnetic fields
generated in the second order vector modes have the amplitude as
$B\approx 10^{-26}$ G around recombination at $k\approx 0.01$Mpc$^{-1}$.  Comparing the amplitudes,
the magnetic fields from the homogeneous solution considered in this
paper will dominate the
density perturbation induced magnetic fields if $r_v \gtrsim 10^{-8}$.

{ Another way to regularize the cosmological vector mode in the
early universe will be a modification of the vector sector of
gravity. An interesting example is Einstein-Aether theory which has been
recently investigated with cosmological perturbations (see, e.g.,
\cite{2010JCAP...07..010A,2011arXiv1103.2197N}). In this class of models
the vector metric now becomes a dynamical variable, and the Aether field
can induce another growing mode in the vector cosmological perturbations. It
is expected that a sizable amount of vector type velocities in the
fluids is induced \cite{2011arXiv1103.2197N}. The magnetic field
generation from this vector mode will be an interesting subject, but beyond
the scope of this paper.  }

In conclusion, in this
paper it is found that the magnetic fields generated from the primordial
vector mode at recombination are allowed at most $B\lesssim 10^{-21}$ G
at $k\approx 0.1$ Mpc$^{-1}$. The upper bound comes from the upper bound
on the primordial vector mode amplitude obtained from the seven-year WMAP
data. This field amplitude is too small to be directly observed
through Faraday effects on the distant radio sources and/or delays of
high energy photons from gamma-ray bursts or blazars, however it may be
large enough for the fields to be a seed for galactic magnetic fields
observed today.

\acknowledgments 
One of the authors (K.I.) would like to thank
T. K. Suzuki and D. G. Yamazaki for helpful comments and useful
discussions.  This work has been supported in part by Grant-in-Aid for
Scientific Research Nos. 22012004 (K.I), 23740179 (K.T.), and 22340056
(N.S.) of the Ministry of Education, Sports, Science and Technology
(MEXT) of Japan, and also supported by Grant-in-Aid for the Global
Center of Excellence program at Nagoya University "Quest for Fundamental
Principles in the Universe: from Particles to the Solar System and the
Cosmos" from the MEXT of Japan. This research has also been supported in
part by World Premier International Research Center Initiative, MEXT,
Japan.


\bibliography{FirstVec}

\end{document}